# The Statistical Coherence-based Theory of Robust Recovery of Sparsest Overcomplete Representation

Lianlin Li

Department of Petroleum Engineering, Texas A&M University

College Station, TX, USA, 77843

**Abstract**

The recovery of sparsest overcomplete representation has recently attracted intensive research activities owe to its important potential in the many applied fields such as signal processing, medical imaging, communication, and so on. This problem can be stated in the following, i.e., to seek for the sparse coefficient vector $\mathbf{x}$ of the given noisy observation $\mathbf{y}$ over a redundant dictionary $\mathbf{D}$ such that $\mathbf{y} = \mathbf{Dx} + \mathbf{n}$, where $\mathbf{n}$ is the corrupted error. Elad et al. made the worst-case result, which shows the condition of stable recovery of sparest overcomplete representation over $\mathbf{D}$ is $\|\mathbf{x}\|_0 \leq \frac{1}{2}\left(\mu^{-1} + 1\right)$ where $\mu(\mathbf{D}) = \max_{i \neq j} \left|\langle \mathbf{d}_i, \mathbf{d}_j \rangle\right|$. Although it's of easy operation for any given matrix, this result can't provide us realistic guide in many cases. On the other hand, most of popular analysis on the sparse reconstruction relies heavily on the so-called RIP (Restricted Isometric Property) for matrices developed by Candes et al., which is usually very difficult or impossible to be justified for a given measurement matrix.

In this article, we introduced a simple and efficient way of determining the ability of given D used to recover the sparse signal based on the statistical analysis of coherence coefficients $\mu_{i,j}$, where $\mu_{i,j}$ is the coherence coefficients between any two different columns of given measurement matrix $\mathbf{D}$. The key mechanism behind proposed paradigm is the analysis of statistical distribution (the mean and covariance) of $\mu_{i,j}$. We proved that if the resulting mean of $\mu_{i,j}$ are zero, and their covariance are as small as possible, one can faithfully recover approximately sparse signals from a minimal number of noisy measurements with overwhelming probability. The resulting theory is not only suitable for almost all models — e.g. Gaussian, frequency measurements—discussed in the literature of compressed sampling, but also provides a framework for new measurement strategies as well.

## I. Introduction

Formally, the problem of interest can be formulated into recovering the *N*-dimensional sparse signal from the corrupted *n*-dimensional observations

$$\mathbf{y} = \mathbf{D}\mathbf{x} + \mathbf{n} \qquad (1)$$

where $\mathbf{D} \in \mathbb{R}^{n \times N}$ ($n < N$) whose columns have unit Euclidean norm are the general highly underdetermined measurement matrix or over-complete basis, $\mathbf{n} \in \mathbb{R}^{n \times 1}$ is measurement error and assumed to follow the Gaussian distribution. The vector $\mathbf{x}$ is assumed to be (approximately) sparse, i.e., its main energy (in terms of the sum of absolute values) is concentrated in only a few entries. Is it possible to faithfully recover a nearly sparse signal $\mathbf{x}$, one which is well approximated by its $k$ largest entries, from incomplete (even highly incomplete) observation $\mathbf{y}$ corrupted by noise? Finding the sparse solution of equation (1) has get more and more interesting since the last two decades, and has played important roles in many applied fields such as medical imaging, signal/imaging processing, and others. Despite considerable progress in the relevant fields, some important questions are still open. We discuss this problem that have both a theoretical and practical appeal in this paper.

The early paper [2-6] triggered a massive amount of research by showing that the $k$-sparse solution $\mathbf{x}$ satisfying the inequality of $\|\mathbf{x}\|_0 \leq \frac{1}{2}\left(\frac{1}{\mu}+1\right)$ is the unique solution to equation (1), and can be efficiently determined with matching pursuit, basis pursuit, and other algorithms, where $\mu(\mathbf{D}) = \max_{i \neq j} |\langle \mathbf{d}_i, \mathbf{d}_j \rangle|$ is the coherence coefficient of $\mathbf{D}$. Though this criterion can be rather easily calculated for any given measurement matrix $\mathbf{D}$, this kind of analysis belongs to so-called worst-case or overwhelming pessimistic result. Actually, the empirical results showed that it is just a simple but limited portrait of the ability of concrete algorithms to find sparse solutions and near-solutions, and far beyond the coverage of the above-described theoretical bound. For example, we used the MATLAB function *randn*(100,500) to generate the measurement matrix $\mathbf{D}$ with size of 100 by 500. The value of coherence $\mu(\mathbf{D})$ in this experiment is about 0.4383, accordingly, the so that only for cardinalities lower than $\|\mathbf{x}\|_0 \leq \frac{1}{2}\left(\frac{1}{\mu(\mathbf{D})}+1\right) \approx 2$ pursuit methods are guaranteed to succeed. However, the empirical results even with pursuit algorithms such as iterative hard threshold algorithm [1, 11], CoSaMP [12], and others show that the solution with cardinalities low than about 15 can be recovered with the overwhelming probability. Actually, many researcher theoretically shows that many exiting random-like matrices has much stronger ability of recovering sparse signal [8-10], for example, for the measurement matrix whose entries come from the i.i.d. Gaussian variable, the *k*-sparse *N*-dimensional signal can be exactly recovered with the measurements with the order of O(*klog(N/k)*) by solving l1-norm constraint convex optimization problem. On the contrary, at present the popular way of addressing these types of problems in the field is by means of the restricted isometry property (RIP), which reveals that the *k*-sparse *N*-dimensional signal can be exactly or stably reconstructed by solving the l1-norm constraint convex optimization under the condition of $\delta_{2k} < \sqrt{2}-1$, where $\delta$ is the so-called RIP constant [10]. Though the elegant theorem, the trouble here is that it is unknown whether or not this property holds for given measurement matrix $\mathbf{D}$, therefore the restricted isometry machinery does not directly apply in this setting. Of course, several efforts have been made to break up this bottleneck, e.g., [13].

This paper develops new theory to recover signals that are approximately sparse in some general (i.e., basis, frame, over-complete, or incomplete) dictionary but corrupted by a combination of measurement noise and interference having a sparse representation in a second general dictionary. The basic goal of this paper is to provide an efficient way of telling the ability of reconstructing k-sparse N-dimensional signal for given measurement matrix **D** in the respect of statistical analysis. To address this problem, this paper introduces a simple and very general way based on the analysis of coherence coefficients of measurement matrix. In this theory, the mechanism behind the proposed approach consists of the following points, namely, firstly, to simply calculate the coherence coefficients between two different columns of given measurement matrix, secondly, to plot the histogram (equivalently, the probability) of obtained coherence coefficients, finally, to estimate the mean and covariance of histogram. The assumption carried out in our theory is that the derived coherent coefficients are i.i.d random number. We proved that if the resulting mean of coherent coefficients are zero, and their covariance are as small as possible, one can faithfully recover approximately sparse signals from a minimal number of noisy measurements. The resulting theory is suitable for almost all models — e.g. Gaussian, frequency measurements —discussed in the literature, but also provides a framework for new measurement strategies as well. The novelty is that our recovery results do not require the restricted isometry property— they make use of a much weaker notion —for the signal.

The rest of this paper is organized as following. In section II, the statistical coherence-based RIP will be studied in detail, which shows that the ability of given **D** used to recover the sparse signal can be efficiently judged by simply analyzing the histogram of computed coherence coefficients. The results show that for the **D** whose entries are generated by i.i.d. zero-mean Gaussian random number, the required observation with number of O(8k) can be used to recover the *k*-sparse *N*-dimensional signal with overwhelming probability. Afterwards, in section III we will discuss the application of proposed statistical coherence-based RIP's in the l1-norm-regularized k-sparse (i.e., only the values of *k* entries are nonzero) signal reconstruction or sparest representation over redundant dictionary. Specifically, we derived the corresponding recovery conditions that guarantee their stability, and discussed the separation of two different signals which are sparse in two different dictionaries or frames. Finally, some conclusions are summarized in section IV.

## II. Statistical Coherence-based RIP

Concerned with the reconstruction of high-dimensional sparse signal from far fewer observations, the key or starting point is the analysis of upper/lower bound of $\frac{\|\mathbf{D}\mathbf{x}_k\|_2^2}{\|\mathbf{x}_k\|_2^2}$ for any non-zero *k*-sparse *N*-dimensional vector $\mathbf{x}_k$. In this section, we will provide a new insight into $\frac{\|\mathbf{D}\mathbf{x}_k\|_2^2}{\|\mathbf{x}_k\|_2^2}$ from the viewpoint of statistical analysis of the coherence coefficient $\mu_{i,j} = \langle \mathbf{d}_i, \mathbf{d}_j \rangle$, where $\mathbf{d}_i$ is the *i*-th column of $\mathbf{D}$. Using obtained results as foundations, we arrived at a rather simple and efficient way of determining the ability of reconstructing the sparse signal for a given matrix $\mathbf{D}$ in the sense of probability.

Firstly, we simply figure out our motivation of proposed methodology. According to the well-known Gershgorin's theorem, one can readily get the following estimation about the bound of $\frac{\|\mathbf{D}\mathbf{x}_k\|_2^2}{\|\mathbf{x}_k\|_2^2}$, i.e.,

$$1 - \mu(k-1) \le \frac{\|\mathbf{D}\mathbf{x}_k\|_2^2}{\|\mathbf{x}_k\|_2^2} \le 1 + \mu(k-1) \qquad (2)$$

where $\mathbf{x}_k$ is the $k$-sparse $N$-dimensional signal, which means that only $k$ ones among the $N$ components of $\mathbf{x}_k$ are nonzero. In the following we will provide the heuristic insight into equation (2) in regard of statistical analysis. For most of measurement matrix $\mathbf{D}$ used in the literatures of compressed sensing, the coherence of $\mu_{i,j} = \langle \mathbf{d}_i, \mathbf{d}_j \rangle$ between any two columns $\mathbf{d}_i$ and $\mathbf{d}_j (i \ne j)$ from $\mathbf{D}$ can be proved to be the i.i.d. zero-mean random variable. Actually, a special case concerns the case where $\mathbf{D}$ also known in the field as the Gaussian measurement ensemble. Another special case is the binary measurement ensemble where the entries of $\mathbf{D}$ are symmetric Bernoulli variables taking on the values $\pm 1$. Then, the statistical version of equation (2) can be heuristically reformulated into

$$0 < 1 - \mu_{i,j}(k-1) \le \frac{\|\mathbf{D}\mathbf{x}_k\|_2^2}{\|\mathbf{x}_k\|_2^2} \le 1 + \mu_{i,j}(k-1) < \infty \qquad (3)$$

Now the task left for us is to determine the probability condition of equation (3) hold; equivalently, the probability condition of $|\mu_{i,j}(k-1)| < 1$. To end this, by means of analytical or numerical estimation we try to model the probability distribution of $\mu_{i,j}$ which is approximately established at least for the measurement matrices appeared in the literatures of compressed sampling. For example, for the matrix whose entries are drawn from i.i.d Gaussian distribution of $\mathcal{N}(0, 1/n)$, the probability of $\mu_{i,j}$ follows $\mu_r \sim \mathcal{N}(0, \sigma_\mu^2)$ with $\sigma_\mu^2 = 1/n$. Now, we can arrive at the probability condition of

$|\mu_r(k-1)| < 1$ as

$$\Pr(|\mu_r(k-1)| < 1) = \Pr\left(|\mu_r| < \frac{1}{k-1}\right)$$
$$= \sqrt{\frac{2}{\pi\sigma_\mu^2}} \int_0^{\frac{1}{k-1}} \exp\left(-\frac{x^2}{2\sigma_\mu^2}\right) dx \qquad (4)$$
$$\geq 1 - \exp\left(-\frac{1}{2(k-1)^2 \sigma_\mu^2}\right)$$

From equation (5) we can get the basic conclusion that $\Pr(|\mu_r(k-1)| < 1) > 0.8647$ if $k \leq 1 + \frac{1}{2\sigma_\mu}$; consequently, if $k \leq \frac{1}{2}\left(1 + \frac{1}{2\sigma_\mu}\right)$ the sparsest solution to equation (1) is unique with overwhelming probability. Compared with the well-known condition of $k \leq \frac{1}{2}\left(1 + \frac{1}{\mu}\right)$ from the worst-case analysis by Elad et al., our heuristic result will be slightly stronger in some cases. For example, we use the function of $\frac{1}{\sqrt{200}} randn(200, 400)$ in MATLAB to generate $\mathbf{D}$ with size of 200 by 400. Then the resulting values of $\mu$ and $2\sigma_\mu = 2\sqrt{n}$ with $n = 200$ are 0.3124 and 0.1414, respectively; consequently, this paper proposed the condition of unique sparsest solution to equation (1) are $k \leq 4$ with overwhelming probability instead of $k \leq 2$.

Different from heuristic investigation discussed above, in the following we will present the details of two strict approaches to the statistical analysis on the coherence-based RIP or the upper/lower bound of $\frac{\|\mathbf{D}\mathbf{x}_k\|_2^2}{\|\mathbf{x}_k\|_2^2}$.

*APPROACH 1.*

Our starting point is to carry out the analysis of $\|\mathbf{D}\mathbf{x}_k\|_2^2 - \|\mathbf{x}_k\|_2^2$ through the central limitation theorem, in particular,

$$\|\mathbf{D}\mathbf{x}_k\|_2^2 - \|\mathbf{x}_k\|_2^2 = \left\|\sum_{i\in\mathrm{supp}(\mathbf{x}_k)}\mathbf{d}_i x_i\right\|_2^2 - \|\mathbf{x}\|_2^2 = \sum_{i\in\mathrm{supp}(\mathbf{x}_k)}\sum_{i\in\mathrm{supp}(\mathbf{x}_k),i\neq j}\mu_{i,j}x_i x_j \tag{5}$$

Taking the assumption made previously that $\mu_{i,j}$ is i.i.d. zero-mean random number with variance $\sigma_\mu^2$ which can derived by analytical or numerical methods of plotting the histogram of $\mu_{i,j}$, one can straightforward get the following conclusion using the well-known Bernstein inequality [14], in particular,

$$\Pr\left(\left|\sum_{i\in\mathrm{supp}(\mathbf{x}_k)}\sum_{i\in\mathrm{supp}(\mathbf{x}_k),i\neq j}\mu_{i,j}x_i x_j\right| > t\right) \leq 2\exp\left(-\frac{t^2}{2\sigma_\mu^2(k-1)\|\mathbf{x}\|_2^4}\right) \tag{6}$$

which means $\left|\|\mathbf{D}\mathbf{x}_k\|_2^2 - \|\mathbf{x}_k\|_2^2\right| \leq 2\sigma_\mu\sqrt{k-1}\|\mathbf{x}_k\|_2^2$ holds on with overwhelming probability. Now we will summary this result in the following theorem, i.e.,

*Theorem 1*

Introducing the notation of $\mu_{i,j} = \langle \mathbf{d}_i, \mathbf{d}_j \rangle$ where $\mathbf{d}_i$ is the $i$-th column of $\mathbf{D}$. Assuming that $\mu_{i,j}$ is i.i.d. zero-mean random number with variance $\sigma_\mu^2$, then for any $k$-sparse vector $\mathbf{x}_k$, the following conclusion holds on

$$1 - 2\sigma_\mu\sqrt{k-1} \leq \frac{\|\mathbf{D}\mathbf{x}_k\|_2^2}{\|\mathbf{x}_k\|_2^2} \leq 1 + 2\sigma_\mu\sqrt{k-1} \tag{7}$$

with overwhelming probability.

*Sketch Proof of theorem 1.*

It is easily proved that the mean and variance of $\|\mathbf{D}\mathbf{x}_k\|_2^2 - \|\mathbf{x}_k\|_2^2$ are

$$E\left(\sum_{i=1}^{k}\sum_{j=1,i\neq j}^{k}\mu_{i,j}x_ix_j\right)=0 \qquad (8)$$

and

$$\begin{aligned}Var\left(\sum_{i=1}^{k}\sum_{j=1,i\neq j}^{k}\mu_{i,j}x_ix_j\right) &= E\left(\sum_{i=1}^{k}\sum_{j=1,i\neq j}^{k}\mu_{i,j}x_ix_j\right)^2 \\ &= \sigma_\mu^2 \sum_{i=1}^{k}\sum_{j=1,i\neq j}^{k}(x_ix_j)^2 \\ &= \sigma_\mu^2\left[\left(\sum_{j=1}^{k}x_i^2\right)^2-\sum_{j=1}^{k}x_i^4\right] \\ &\leq \sigma_\mu^2(k-1)\left(\sum_{j=1}^{k}x_i^2\right)^2 \\ &= \sigma_\mu^2(k-1)\|\mathbf{x}\|_2^4\end{aligned} \qquad (9)$$

respectively. Substituting equations (8) and (9) into the so-called Bernstein theorem [14], we can straightforward finish the proof of theorem 1.

From theorem 1, we can derive one important conclusion that the unique condition of sparsest solution to equation (1) is $k \leq \frac{1}{2}\left(1+\frac{1}{4\sigma_\mu^2}\right)$ with overwhelming probability (Note: for matrix **D** whose entries come from the *i.i.d.* zero-mean Gaussian random number with variance of $1/n$, the resulting condition is $k \leq \frac{1}{2}\left(1+\frac{n}{4}\right)=O(n)$). For **D** generated above, the corresponding condition is $k \leq 25$ with overwhelming probability. Obviously, there is important improvement on the results achieved previously.

*APPROACH 2.*

In the following we will provide alternative the analysis of $\|\mathbf{Dx}_k\|_2^2 - \|\mathbf{x}_k\|_2^2$ via the well-known operator-Bernstein inequality. To do this, we introduce the notation of $\mathbf{D}_k$ denoting the sub-matrix generated by the some $k$ columns of $\mathbf{D}$. Then we have the following theorem about the bound of $\|\mathbf{D}_k^T \mathbf{D}_k - \mathbf{I}\|$, in particular,

*Theorem 2*

Introducing the notation of $\mu_{i,j} = \langle \mathbf{d}_i, \mathbf{d}_j \rangle$ where $\mathbf{d}_i$ is the i-th column of $\mathbf{D}$. Assuming that $\mu_{i,j}$ is *i.i.d.* zero-mean random number with variance $\sigma_\mu^2$, then we have the following conclusion, i.e.,

$$\Pr\left(\|\mathbf{D}_k^T \mathbf{D}_k - \mathbf{I}\| > t\right) \leq \frac{k(k-1)}{2} \exp\left(-2\frac{t^2}{4k(k-1)\sigma_\mu^2}\right)$$

*Sketch proof of theorem 2*

To proof theorem 2, the matrix of $\mathbf{D}_k^T \mathbf{D}_k - \mathbf{I}$ can be reformulated into

$$\mathbf{D}_k^T \mathbf{D}_k - \mathbf{I} = \begin{bmatrix} \mathbf{d}_1^T \\ \vdots \\ \mathbf{d}_k^T \end{bmatrix} [\mathbf{d}_1, \ldots, \mathbf{d}_k] - \mathbf{I}$$

$$= \begin{bmatrix} 0 & \mu_{12} & & & \\ \mu_{12} & 0 & & \mu_{i,j} & \\ & & \ddots & & \\ \mu_{i,j} & & & 0 & \\ & & & & 0 \end{bmatrix} \quad (10)$$

$$= \sum_{i=1}^{k} \sum_{j=i+1}^{k} \mu_{i,j} \mathbf{\Delta}_{i,j}$$

where $\mathbf{\Delta}_{i,j}$ is the matrix whose entry is unit at the locations of $(i,j)$ or $(j,i)$, and zero otherwise. It is readily proved that $E(\mu_{i,j} \mathbf{\Delta}_{i,j}) = \mathbf{0}$ and $\|E(\mu_{i,j}^2 \mathbf{\Delta}_{i,j}^2)\| \leq \sigma_\mu^2$. Finally, using operator-Bernstein theorem can straightforward yield the following conclusion, i.e.,

$$\Pr\left(\left\|\mathbf{D}_k^T \mathbf{D}_k - \mathbf{I}\right\| > t\right) \leq \frac{k(k-1)}{2} \exp\left(-2 \frac{t^2}{4k(k-1)\sigma_\mu^2}\right) \quad (11)$$

which finishes the proof of theorem 2.

Theorem 2 tells us that the sparest solution to equation (1) is unique with overwhelming probability if $k \leq \frac{1}{4\sigma_\mu}$; in other words, the inequality of

$$\left|\|\mathbf{D}\mathbf{x}_k\|_2^2 - \|\mathbf{x}_k\|_2^2\right| \leq 2\sigma_\mu \sqrt{k(k-1)} \|\mathbf{x}_k\|_2^2$$

holds on with overwhelming probability. Again, for the matrix **D** generated previously, in particular, i.i.d. zero-mean Gaussian random number with variance of $1/n$, the corresponding condition becomes $k \leq \frac{\sqrt{n}}{4}$. The estimate we have just seen is not isolated and the real purpose of this section is to develop a theory of compressive sensing which is both as simple and as general as possible.

## III. Application in the stable reconstruction of sparse signal

Based on the statistical coherence-based conclusions (i.e., theorem 1 and theorem 2) derived above, in this section we will discuss its application in the l1-norm-regularized $k$-sparse (i.e., only the values of $k$ entries are nonzero) signal reconstruction or sparest representation over redundant dictionary. Mathematically, this problem can be formulated into

$$\min \|\mathbf{x}\|_1 \quad s.t., \|\mathbf{y} - \mathbf{D}\mathbf{x}\|_2 \leq \varepsilon \quad (12)$$

where $\varepsilon$ is the upper bound of measurement error. Now our first task is to derive the corresponding recovery conditions that guarantee their stability. Afterwards, we will further turn to the discussion of separation of two different features which are sparse in two distinct dictionaries or frames, which can be generalized into the separation of several different features. Mathematically, this problem can be formulated into

$$\min \|\mathbf{x}\|_1 + \|\mathbf{e}\|_1 \qquad s.t. \|\mathbf{y} - \mathbf{Dx} - \mathbf{Be}\|_2 \leq \varepsilon \tag{13}$$

It is noted that solving problem (13) is appealing in several applied applications as listed in the following two examples, i.e.,

*(a) the robust sparse signal reconstruction, i.e.,*

$$\mathbf{y} = \mathbf{Dx} + \mathbf{e} + \mathbf{n} \tag{14}$$

where the measurement error consists of two terms of $\mathbf{e}$ and $\mathbf{n}$, $\mathbf{n}$ is the usual Gaussian noise with small value while $\mathbf{e}$ is the sparse vector of measurement error probably due to measurement mismatch or/and random malfunctioning of the instrument for measuring $\mathbf{y}$, and other possible reasons. It is noted that there is no other knowledge about $\mathbf{e}$.

*(b) signal separation, i.e.,*

$$\mathbf{y} = \mathbf{Dx} + \mathbf{Be} \tag{15}$$

The setting (15) allows us to study the signal separation, i.e., the separation of two distinct features $\mathbf{Dx}$ and $\mathbf{Be}$ from the noisy observation $\mathbf{y}$ corrupted by white noise $\mathbf{n}$. It should be pointed out that in equation (15) $\mathbf{Dx}$ and $\mathbf{Be}$ are two desirable features whose respective coefficients in $\mathbf{D}$ and $\mathbf{B}$ are sparse vectors, i.e., $\|\mathbf{x}\|_0 \leq n_x \ll N_x$ and $\|\mathbf{e}\|_0 \leq n_e \ll N_e$ while $\dim(\mathbf{x}) = N_x$ and $\dim(\mathbf{e}) = N_e$. Signal separation amounts to simultaneously recovering the vectors $\mathbf{x}$ and $\mathbf{e}$ from the noisy measurement $\mathbf{y}$ followed by computation of the individual signal features **Ax** and **Be**. Still, there is no any knowledge about **x** and **e** is available, except for the fact that each vector exhibits an approximately sparse representation in both **D** and **B**.

Firstly, we consider the application of proposed theorem 1 (or theorem 2) for stably solving the problem represented by equation (12), which has been summarized in theorem 3, in particular,

*Theorem 3*

Introducing the notation of $\|\mathbf{x}\|_0 \leq k$ and $\mu_{i,j} = \langle \mathbf{d}_i, \mathbf{d}_j \rangle$ where $\mathbf{d}_i$ is the i-th column of **D**. Assuming that $\mu_{i,j}$ is *i.i.d.* zero-mean random number with variance $\sigma_\mu^2$, then the stable sparest solution to (12) holds on with overwhelming probability if $k \leq 1 + \frac{1}{9}\sigma_\mu^{-2}$.

*Sketch proof of theorem 3*

Assuming that $\hat{\mathbf{x}}$ is the output of solving eq. (12) while $\mathbf{x}_0$ is the true one. Introducing the notation of $\mathbf{h} = \hat{\mathbf{x}} - \mathbf{x}$, then one has the estimation of $\|\mathbf{Dh}\|_2^2 \leq 4\varepsilon^2$ along the standard operation in the literatures [10, 15]. At the same time, we can derive the low bound of $\|\mathbf{Dh}\|_2^2$ as following, i.e.,

$$\begin{aligned}
\|\mathbf{Dh}\|_2^2 &= \mathbf{h}^T \mathbf{D}^T \mathbf{Dh} \\
&= \sum_{k,l} \left(\mathbf{h}^T\right)_k \mathbf{d}_k^T \mathbf{d}_l \left(\mathbf{h}\right)_l \\
&= \sum_k \left|\left(\mathbf{h}^T\right)_k\right|^2 + \sum_{k \neq l} \left(\mathbf{h}^T\right)_k \mathbf{d}_k^T \mathbf{d}_l \left(\mathbf{h}\right)_l \\
&\geq \|\mathbf{h}\|_2^2 - \mu \sum_{k \neq l} \left|\left(\mathbf{h}^T\right)_k \left(\mathbf{h}\right)_l\right| \\
&= \|\mathbf{h}\|_2^2 + \mu \sum_k \left|\left(\mathbf{h}^T\right)_k\right|^2 - \mu \sum_{k,l} \left|\left(\mathbf{h}^T\right)_k \left(\mathbf{h}\right)_l\right| \\
&= (1+\mu) \|\mathbf{h}\|_2^2 - \mu \|\mathbf{h}\|_1^2
\end{aligned} \quad (16)$$

where $\left(\mathbf{h}\right)_l$ means the $l$th element of vector $\mathbf{h}$. Combing equation (16) with $\|\mathbf{Dh}\|_2^2 \leq 4\varepsilon^2$ readily gives us the upper bound of $\|\mathbf{h}\|_2^2$ as

$$\|\mathbf{h}\|_2^2 \leq \frac{4\varepsilon^2 + \mu \|\mathbf{h}\|_1^2}{1+\mu} \quad (17)$$

Taking the fact of $\|\mathbf{h}\|_1 \leq 2\|\mathbf{h}_T\|_1 + e_0$ together with $e_0 = 2\|\mathbf{x}_{T^c}\|_1$ into account [10,15], from equation (17) we can readily arrive at the low bound of $\|\mathbf{Dh}\|_2^2$ as

$$\begin{aligned}
\|\mathbf{h}\|_2 &\leq \frac{2\varepsilon + \sqrt{\mu}\|\mathbf{h}\|_1}{\sqrt{1+\mu}} \\
&\leq \frac{2\varepsilon + \sqrt{\mu}\left(2\|\mathbf{h}_T\|_1 + e_0\right)}{\sqrt{1+\mu}} \\
&\leq \frac{2\varepsilon + \sqrt{\mu}\left(2\sqrt{k}\|\mathbf{h}_T\|_2 + e_0\right)}{\sqrt{1+\mu}}
\end{aligned} \quad (18)$$

where $\mathbf{h}_T$ denotes the sub-vector of $\mathbf{h}$ at the domain $T$, and the subscription of $T$ is the support corresponding to the nonzero coefficients of $\mathbf{x}$.

On the other hand, we can estimate the upper bound of $\|\mathbf{h}_T\|_2$ by the analysis of $\left|\mathbf{h}^T \mathbf{D}^T \mathbf{Dh}_T\right|$ still along the standard process carried out in the most of publications, in particular, the estimation of

$$\begin{aligned}
\left| \mathbf{h}^T \mathbf{D}^T \mathbf{D} \mathbf{h}_T \right| &= \left| \left( \mathbf{h}_T + \mathbf{h}_{T^c} \right)^T \mathbf{D}^T \mathbf{D} \mathbf{h}_T \right| \\
&\geq \left\| \mathbf{D} \mathbf{h}_T \right\|_2^2 - \left| \left( \mathbf{h}_{T^c} \right)^T \mathbf{D}^T \mathbf{D} \mathbf{h}_T \right| \\
&\geq \left( 1 - g_k \right) \left\| \mathbf{h}_T \right\|_2^2 - \sum_{j \geq 1} \left| \mathbf{h}_{T_j} \mathbf{D}^T \mathbf{D} \mathbf{h}_T \right| \\
&\geq \left( 1 - g_k \right) \left\| \mathbf{h}_T \right\|_2^2 - \sigma_\mu \left\| \mathbf{h}_T \right\|_2 \left\| \mathbf{h}_{T^c} \right\|_1 \\
&\geq \left( 1 - g_k \right) \left\| \mathbf{h}_T \right\|_2^2 - \sigma_\mu \left\| \mathbf{h}_T \right\|_2 \left( \left\| \mathbf{h}_T \right\|_1 + e_0 \right) \\
&\geq \left( 1 - g_k \right) \left\| \mathbf{h}_T \right\|_2^2 - \sigma_\mu \left\| \mathbf{h}_T \right\|_2 \left( \sqrt{k} \left\| \mathbf{h}_T \right\|_2 + e_0 \right)
\end{aligned}$$

will hold on with overwhelming probability, where $g_k = 2\sigma_\mu \sqrt{k-1}$. Further using the inequality of $2\varepsilon \sqrt{1+\sigma_k} \left\| \mathbf{h}_T \right\|_2 \geq \left| \mathbf{h}^T \mathbf{D}^T \mathbf{D} \mathbf{h}_T \right|$ in above result, we can get the upper bound of $\left\| \mathbf{h}_T \right\|_2$ as

$$\left\| \mathbf{h}_T \right\|_2 \leq \frac{2\varepsilon \sqrt{1+g_k} + \sigma_\mu e_0}{1 - g_k - \sqrt{k} \sigma_\mu}. \tag{19}$$

Combing equations (18) and (19), the final estimation of upper bound of $\left\| \mathbf{h} \right\|_2$ can be arrived at

$$\begin{aligned}
\left\| \mathbf{h} \right\|_2 &\leq \frac{2\varepsilon + \sqrt{\mu} \left( 2\sqrt{k} \dfrac{2\varepsilon \sqrt{1+g_k} + \sigma_\mu e_0}{1 - g_k - \sqrt{k} \sigma_\mu} + e_0 \right)}{\sqrt{1+\mu}} \\
&\equiv \frac{\varepsilon c_1 + c_2 e_0}{1 - g_k - \sqrt{k} \sigma_\mu}
\end{aligned} \tag{20}$$

where $c_1$ and $c_2$ are two constants dependent on $k$, $\mu$ and $\sigma_\mu$.

From equation (20), we can get the basic requirement of stable solution to (12) is

$$1 - \sigma_k - \sqrt{k} \sigma_\mu = 1 - 2\sigma_\mu \sqrt{k-1} - \sigma_\mu \sqrt{k} \geq 0 \tag{21}$$

Consequently,

$$k \leq 1 + \frac{1}{9} \sigma_\mu^{-2} \tag{22}$$

Now the proof of theorem 3 is finished.

Theorem 3 shows that the sparest solution to equation (1) is stably obtained by solving the l1-norm regularized optimization problem of (12) with overwhelming probability if $k \leq 1 + \frac{1}{9}\sigma_\mu^{-2}$. Again, for the matrix $\mathbf{D}$ whose entries are come from i.i.d. zero-mean Gaussian random number with variance of $1/n$, the corresponding condition becomes $k \leq 1 + \frac{n}{9}$.

Now we turn to discuss the separation of two distinct features which are sparse in two different dictionaries of $\mathbf{D}$ and $\mathbf{B}$. Mathematically, the basic goal is to get the condition of guaranteeing the stable solution to (13). It is noted that there is no any knowledge about $\mathbf{x}$ and $\mathbf{e}$ is available, except for the fact that each vector exhibits an approximately a sparse representation in both $\mathbf{D}$ and $\mathbf{B}$. To address this problem, similar as requirement by theorem 3, the first thing for us is to get the statistical coherence-based RIP characteristic for joint sparse dictionary case, in particular, to get the upper and low bound of $\|\mathbf{Dx} + \mathbf{Be}\|_2^2$ for any sparse vectors of $\mathbf{x}$ and $\mathbf{e}$, where $\|\mathbf{x}\|_0 \leq n_x$ and $\|\mathbf{e}\|_0 \leq n_e$. For notable convenience, the notations of $\tilde{\mathbf{D}} \equiv [\mathbf{D}, \mathbf{B}]$ and $\tilde{\mathbf{x}} = [\mathbf{x}^T, \mathbf{e}^T]^T$ are introduced.

To obtain the estimation of upper and lower bound of $\|\mathbf{Dx} + \mathbf{Be}\|_2^2$, we firstly use the following identical equation,

$$\|\tilde{\mathbf{D}}\tilde{\mathbf{x}}\|_2^2 = \|\mathbf{Dx}\|_2^2 + \|\mathbf{Be}\|_2^2 + 2\langle \mathbf{Dx}, \mathbf{Be} \rangle \qquad (23)$$

Assuming that the dictionaries of $\mathbf{D}$ and $\mathbf{B}$ obey their own statistical coherence based RIP, i.e.,

$$(1-g_x)\|\mathbf{x}\|_2^2 \leq \|\mathbf{Dx}\|_2^2 \leq (1+g_x)\|\mathbf{x}\| \tag{24}$$

and

$$(1-g_e)\|\mathbf{e}\|_2^2 \leq \|\mathbf{Be}\|_2^2 \leq (1+g_e)\|\mathbf{e}\|_2^2 \tag{25}$$

where $g_x = 2\sigma_{D,\mu}\sqrt{n_x-1}$, $g_e = 2\sigma_{D,e}\sqrt{n_e-1}$, $\sigma_{D,\mu}$ and $\sigma_{D,e}$ the square coherence covariance associated with $\mathbf{D}$ and $\mathbf{B}$, respectively. To estimate the bound of $|\langle \mathbf{Ax}, \mathbf{Be}\rangle|$, the following conclusion can be readily obtained

$$|\langle \mathbf{Dx}, \mathbf{Be}\rangle| = \left|\sum_i \sum_j \langle \mathbf{d}_i, \mathbf{b}_j\rangle x_i e_j\right| \leq \sum_i \sum_j |\langle \mathbf{d}_i, \mathbf{b}_j\rangle||x_i e_j| \tag{26}$$

If $\mu_m = \max_{i,j} |\langle \mathbf{d}_i, \mathbf{b}_j\rangle|$, then one has [15]

$$\begin{aligned}
2|\langle \mathbf{Dx}, \mathbf{Be}\rangle| &\leq 2\sum_i \sum_j |\langle \mathbf{d}_i, \mathbf{b}_j\rangle||x_i e_j| \\
&\leq 2\mu_m \|\mathbf{x}\|_1 \|\mathbf{e}\|_1 \\
&\leq 2\mu_m \sqrt{n_x n_e} \|\mathbf{x}\|_2 \|\mathbf{e}\|_2 \\
&\leq \mu_m \sqrt{n_x n_e} \left(\|\mathbf{x}\|_2^2 + \|\mathbf{e}\|_2^2\right) \\
&= \mu_m \sqrt{n_x n_e} \|\tilde{\mathbf{x}}\|_2^2
\end{aligned} \tag{27}$$

Consequently, from equations (23), (24), (25) and (27) one has the coherence-based RIP similar as shown in theorem 1 and theorem 2

$$\left(1-\max(g_x, g_e) - \mu_m\sqrt{n_x n_e}\right)\|\tilde{\mathbf{x}}\|_2^2 \leq \|\tilde{\mathbf{D}}\tilde{\mathbf{x}}\|_2^2 \leq \left(1+\max(g_x, g_e) + \mu_m\sqrt{n_x n_e}\right)\|\tilde{\mathbf{x}}\|_2^2 \tag{28}$$

Actually this conclusion represented by equation (27) is hold on based on the worst-case analysis. Different from this, here we provide alternative investigation, i.e., the statistical coherence-based analysis. To do this, we take the assumption that the value of $\langle \mathbf{d}_i, \mathbf{b}_j\rangle$ is to be the i.i.d. zero-mean random number with variance of $\sigma_{\mu m}^2 \equiv \mathrm{var}(\langle \mathbf{d}_i, \mathbf{b}_j\rangle)$. Therefore, the inequality of

$$\left|\langle \mathbf{A}\mathbf{x}, \mathbf{B}\mathbf{e}\rangle\right| = \left|\sum_i \sum_j \langle \mathbf{a}_i, \mathbf{b}_j\rangle x_i e_j\right| \leq \sum_i \sum_j \sigma_{\mu m} x_i e_j \leq \sigma_{\mu m} \|\mathbf{x}\|_2 \|\mathbf{e}\|_2 \leq \sigma_{\mu m} \|\tilde{\mathbf{x}}\|_2 \qquad (29)$$

holds on with overwhelming probability. Now, combing equations (23),(24),(25)and (29), we can achieve at the result that the statistical coherence-based RIP of

$$\left(1 - g(n_x, n_e)\right)\|\tilde{\mathbf{x}}\|_2^2 \leq \|\tilde{\mathbf{D}}\tilde{\mathbf{x}}\|_2^2 \leq \left(1 + g(n_x, n_e)\right)\|\tilde{\mathbf{x}}\|_2^2 \qquad (30)$$

holds on with overwhelming probability, where $g(n_x, n_e) = \max(g_x, g_e) + \sigma_{\mu m}$. Based on above results, the following theorem used to guarantee the stable solution to (13) can be readily derived.

*Theorem 4*

For the problem described by equation (13), assuming that $\|\mathbf{x}\|_0 \leq n_x \ll N_x$ and $\|\mathbf{e}\|_0 \leq n_e \ll N_e$ while $\dim(\mathbf{x}) = N_x$ and $\dim(\mathbf{e}) = N_e$, and the joint dictionary of $\tilde{\mathbf{D}}$ follows the statistical coherence-based RIP of equation (30) with overwhelming probability. Then, the stable solution to (13) can be guaranteed if

$$g(n_x, n_e) + \sigma_{\mu m}\sqrt{w} < 1 \qquad (31)$$

where $w = n_x + n_e$.

*Sketch proof of theorem 4*

Similar as done in the proof of theorem 3, the estimation of upper bound of $\|\mathbf{h}\|_2$ can be readily achieved as

$$\|\mathbf{h}\|_2 \leq \frac{2\varepsilon + \sqrt{\mu_d}\left(2\sqrt{w}\|\mathbf{h}_T\|_2 + e_0\right)}{\sqrt{1+\mu_d}} \tag{32}$$

where $\mu_d = \max_{i \neq j}\left|\langle \tilde{\mathbf{d}}_i, \tilde{\mathbf{d}}_j \rangle\right|$. On the other hand, the inequality of

$$\begin{aligned}
\left|\mathbf{h}^T \mathbf{D}^T \mathbf{D}\mathbf{h}_T\right| &= \left|\left(\mathbf{h}_T + \mathbf{h}_{T^c}\right)^T \mathbf{D}^T \mathbf{D}\mathbf{h}_T\right| \\
&\geq \|\mathbf{D}\mathbf{h}_T\|_2^2 - \left|\left(\mathbf{h}_{T^c}\right)^T \mathbf{D}^T \mathbf{D}\mathbf{h}_T\right| \\
&\geq \left(1 - g(n_x, n_e)\right)\|\mathbf{h}_T\|_2^2 - \sigma_{\mu m}\|\mathbf{h}_T\|_2 \left(\|\mathbf{h}_T\|_1 + e_0\right) \\
&\geq \left(1 - g(n_x, n_e)\right)\|\mathbf{h}_T\|_2^2 - \sigma_{\mu m}\|\mathbf{h}_T\|_2 \left(\sqrt{w}\|\mathbf{h}_T\|_2 + e_0\right)
\end{aligned} \tag{33}$$

holds on with overwhelming probability. Taking $2\varepsilon\sqrt{1+g(n_x,n_e)}\|\mathbf{h}_T\|_2 \geq \left|\mathbf{h}^T \mathbf{D}^T \mathbf{D}\mathbf{h}_T\right|$ into account, we can get

$$\|\mathbf{h}_T\|_2 \leq \frac{2\varepsilon\sqrt{1+g(n_x, n_e)} + \sigma_{\mu m}\sqrt{w}e_0}{1 - g(n_x, n_e) - \sigma_{\mu m}\sqrt{w}} \tag{34}$$

Substituting equation (34) into (32) yields

$$\|\mathbf{h}\|_2 \leq \frac{c_1 \varepsilon + c_2 e_0}{1 - g(n_x, n_e) - \sigma_{\mu m}\sqrt{w}} \tag{35}$$

where $c_1$ and $c_2$ are two constants dependent on the parameters of $n_x, n_e$ and $\sigma_{\mu m}$. Now the proof of theorem 4 can be obviously closed.

Theorem 4 shows that compared with worst-case results developed in [15], in the sense of probability the proposed result of $g(n_x, n_e) + \sigma_{\mu m}\sqrt{w} < 1$ provide us much stronger and efficient way of judging the ability of separating two different features sparse in different dictionaries, by means of simply calculating the coherent coefficients, and further analyzing their statistical properties. As mentioned above, the proposed approach can be generalized into the separation of multiple distinct features along the same line as used in this paper.

## IV. Conclusion

The recovery of sparsest overcomplete representation has recently attracted intensive research activities owe to its important potential in the many applied fields such as signal process, medical imaging, communication, and so on. The result of coherence based analysis provide the worst-based results, which shows the condition of stable recovery of sparest overcomplete representation is $\|\mathbf{x}\|_0 \leq \frac{1}{2}(\mu^{-1}+1)$ where $\mu(\mathbf{D}) = \max_{i \neq j}|\langle \mathbf{d}_i, \mathbf{d}_j \rangle|$ is the coherence of $\mathbf{D}$. Although it's of easy operation for any given matrix, this result usually can't provide us helpful guide. On the other hand, most of analysis on the sparse reconstruction relies heavily on the so-called RIP (Restricted Isometric Property) for matrices, which is very difficult or impossible to be justified for a given measurement matrix.

In this article, we introduced a easy-operation way based on the statistical analysis of coherence coefficients $\mu_{i,j}$, where $\mu_{i,j}$ is the coherence coefficients between any two different columns of given measurement matrix $\mathbf{D}$. The key mechanism behind proposed methodology is the statistical distribution (the mean and covariance) of $\mu_{i,j}$. We proved that if the resulting mean of coherent coefficients are zero, and their covariance are as small as possible, one can faithfully recover approximately sparse signals from a minimal number of noisy measurements with overwhelming probability by solving the l1-norm regularization convex optimization problem. The assumption made in our theory is the

i.i.d. random distribution of coherence coefficient $\mu_{i,j}$, which holds on for many measurement matrices used in the literatures of compressed sampling. The investigation of general case for dependent random distribution $\mu_{i,j}$ will be the future work.